# The Determinants For User Intention To Adopt Web Based Early Childhood Supplementary Educational Platform

[1]Khoo Han Ni, [2]Ahmad Suhaimi Baharudin, [3*]Kamal Karkonasasi

[123]School of Computer Sciences, Universiti Sains Malaysia (USM), 11800, Penang, Malaysia.

[3]*asasi.kamal@gmail.com

**Abstract**

Education is defined as the fundamental key for the success and development for any given society. Hence, the forming of a strong basic educational foundation is crucial to ensure the children to stay competitive and achieve extraordinary in the changing world. I-zLink Sdn. Bhd. is a partnership company that formed by four individuals with the commitment to develop an early childhood learning solution named i-Future. The system, i-Future is a Supplementary Educational Platform that designs specially for children that fall between the age categories of two to six years which integrate with the latest technology to increase the engagement and flexibility in learning process as well as the effectiveness of study. In order to ensure that the system has a market value, an empirical study was conducted to determine the acceptance level and also the variables that would affect the user's intention to adopt i-Future. The variables used in the study include Perceived Ease of Use (PEU), Perceived Usefulness (PU), System Quality (SQ) and Social Norm (SN). Through the quantitative study, the relationship between these variables and user's intention to adopt i-Future was determined. Those variables which have significant impact on user's intention to adopt i-Future will be used to design i-Future.

## 1. Introduction

In accordance to the United Nations Millennium Development Goals, "education" is believed to be a fundamental key for success and development, which eradicates poverty, disease and improves the health, economic, stability and quality of life for any given society. In fact, education is defined as the process of transmission of knowledge, skills, and values from one generation to another. It also opens the door towards success and development of any given society [1]. Hence, study is done to determine the real problem or the root cause plaguing in the current educational system in order to improve the quality of the students and increase the knowledgeable society for high-income nation. In fact, the foundation development or the nurturing in the earliest years is crucial for cognitive and pre-literacy skills development [2].

Rima Shore, the author for "Rethinking the Brain" also justified that "the development and the experiences in the early years actually can impact the brain architecture as well as the nature and extent of audit capacities" [3]. Therefore, we intend to develop i-Future which is a Web-based Supplementary Educational Platform that specially designed for the children that fall between the age categories of 2 to 6 to "enter to learn, leave to achieve". The aim of i-Future is to build up the pre-literacy and cognitive skills for children during their earliest formative years through the integration of supplementary educational services with the technology solution. In the following section, Literature review has been done to identify the gap in the current education system and the root cause of it. The current methods are explained in section 3 and the proposed research method and the data collection are presented in section 4 and 5 consecutively. The statistical analysis is in section 6 and finally the discussion and conclusion is presented in section 7 and 8 respectively.

## 2. Literature Review

According to the Government Transformation Programmer (GTP) Roadmap, "Malaysia's student outcomes have deteriorated". In fact, 20% of the students are failing to meet the minimum benchmark for TIMSS (Trends in International Mathematics and Science Study) in year 2007 in compare to 5-7% in year 2003. Besides, the report also stated that in year 2008, almost 32,000 students dropped out of school at various stages. A global ranking conducted by the Organisation for Economic Co-operation and Development (OECD) think tank on year 2015. Malaysia is ranked at 52th out of 72 countries. The ranking is established on test scores of 15 year-olds in maths and science [4]. Therefore, education quality in Malaysia is far behind the countries such as Hong Kong, South Korea or even Singapore. Dorothy E. Leidner et. al. criticizes that the current existing educational system is less sufficient to engage the children as well as create greater satisfaction in educational process [5]. Besides, they also criticizes that the teaching method is with little concern whether students can turn the knowledge into cognitive scheme. Their proposed solution is to learn as a group and using multimedia format to engage the students [5]. Malaysia Achieving The Millennium Development Goals: Success and Challenges Report stated, "Some children who perform well in examinations may not necessarily understand fundamental concepts and are therefore unable to apply, for instance, mathematical and science concepts outsides the school or textbook context". The proposed solution in the report is to increase the quality of the educators and participation from the private sector to increase the enrolment for pre-school and primary level [6]. According to Experience and Education analysis done by John Dewey, "the conventional or traditional curriculum always ignore the capabilities and interests of child nature" [7]. The same situation still exists in current education system which include Malaysia. In fact, the current education system is too result-oriented, fixed and fail in breeding critical think.

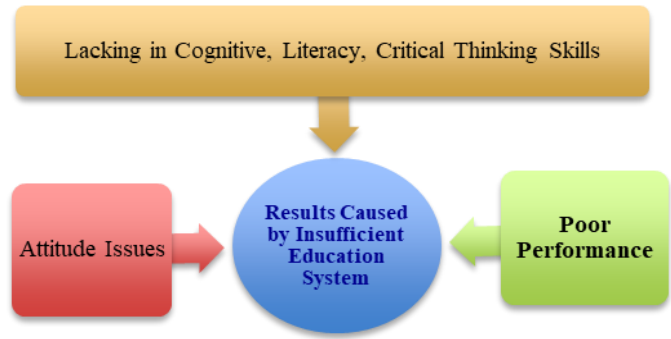

Figure 1. Results Caused by Insufficient Educational System

Figure 1 shows that the current educational system is insufficient. In fact, the insufficient of education system is the root causes that contribute to the issues such as poor attitude, lacking in cognitive, literacy and critical thinking skills as well as poor performance. These challenges are going to affect the development of a nation.

## 3. Current Methods

During the last few decades, the theoretical models are designed with the diffusion of innovation in order to test the acceptance level of user intention to adopt in Information Technology [8] and Information System (IT/IS) [9]. In fact, this research is crucial for the researchers to understand why people resist using the Information Technology, how users will respond to them and what are the critical success factors to improve the user acceptance level in order to design a high market demand product/service [8]. Below are the existing theoretical frameworks that commonly used by the researchers to test the adoption or acceptance level towards the Information Technology and Information System (IT/IS).

### 3.1 Theory Acceptance Model (TAM)

The Technology Acceptance Model (TAM) is originated by Davis et. al derived from Theory of Reasoned Action. In fact, the user intention is influenced by the user's two beliefs about the system: Perceived Usefulness (PU) and Perceived Ease of Use (PEU). From Davis's research finding, it shows that both PU and PEU have the positive effect on user intention to use and PEU has a positive causal effect on PU [8]. Since the establishment of the model, TAM are widely use by research to test the degree of acceptance in various application such as e-services, e-learning, information management system, e-commerce, e-banking, mobile service, mobile learning, e-recruitment system and etc. TAM model is very helpful for forecasting and evaluating user acceptance of Information Technology. However, different system or services require the different additional variables instead of PU and PEU in order to test the specific context so that the research result can provides more useful information which is crucial for the system design/characteristics as "lack of user acceptance has long been an impediment to the success of new information system" [8].

## 3.2 Theory of Planned Behavior (TPB)

The Theory of Planned Behavior which is illustrated in Figure 2, is the conceptual framework that typically use for dealing with the complexity of human social behavior [10]. It always been used to predict and explain particular behaviors in specified contexts. The model is originated by Azjen et. al derived from Theory of Reasoned Action [10]. According to Azjen et. al, human intention to perform behaviors of different kinds could be predicted with high accuracy from the attitudes towards behavior, subjective norm, and perceived behavior control [11].

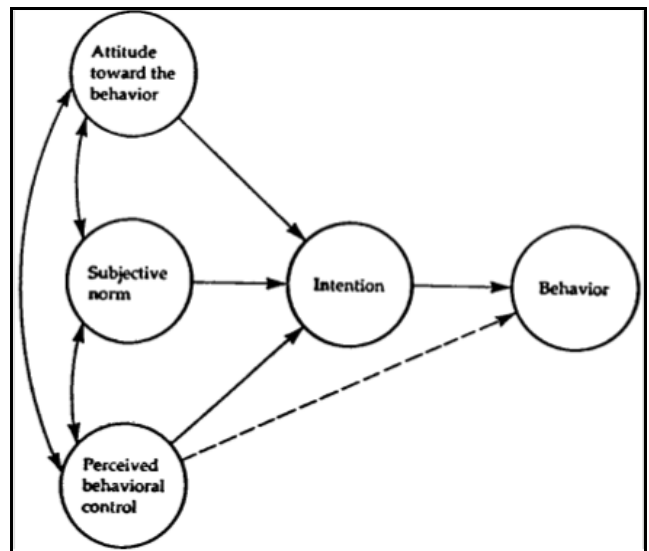

Figure 2. Theory of Planned Behavior [10]

## 3.3 Information System (IS) Success Model

Information System (IS) Success Model as shown in Figure 3, is originated form William H. DeLone and Ephraim R. McLean for conceptualizing and operationalising IS success [12]. According to the model, there are six constructs of Information Success, which include System Quality, Information Quality, Use, User Satisfaction, Individual Impact and Organizational Impact [12]. The system quality is defined as the "desired characteristics of the information system itself which produce the information" whereas information quality is defined as the "desired characteristics such as accuracy, meaningfulness and timeliness".

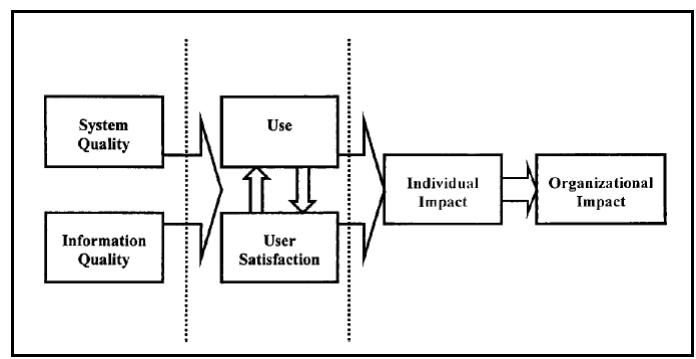

Figure 3. Information System (IS) Success Model [12]

## 4. Proposed Research Method

The hypotheses are define base on the proposed research method shown in Figure 4. The hypotheses are listed as follow.

> **H1**: Perceived Ease to Use of will have positive effect on Behavioral Intention to adopt i- Future in Malaysia.
>
> **H2**: Perceived Usefulness will have positive effect on Behavioral Intention to adopt i-Future in Malaysia.
>
> **H3**: System Quality has the positive effect on Behavioral Intention to adopt i-Future in Malaysia.
>
> **H4**: Social Norm has positive effect on Behavioral Intention to adopt i-Future in Malaysia.

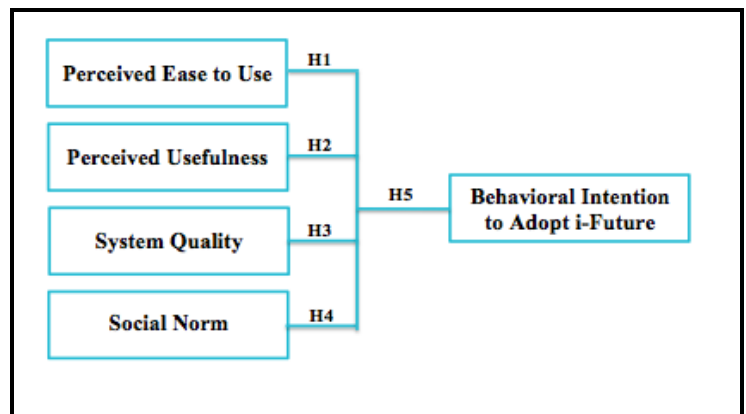

Figure 4: The Proposed Research Model

## 5. Data Collection

A number of attributes are used to design the survey questions. Gender, age, race, marital status, number of children, education level, employment, income, mobile/pc usage and subscription rate are the attributes nominated in the survey. This survey is conducted to the specific target group, who are those married individual. This is because the i-Future is designed for early childhood education and the parents are the target customers. Hence, their acceptance level needs to be tested. There are total of 104 copies of sample data being collected. The respondents are from different background and educational level. In fact, 98.1% of the respondents owned mobile devices or personal computer. Thus, it proof that the Internet penetration in Malaysia is high. However, the subscription level of Malaysian is only 45%. Hence, more promotion and pricing strategy need to be carried out in order to attract more customers.

## 6. Statistical Analysis

a) **Reliability test:** The reliability test is to measure the consistency of the items in each of the variable when given the same measurement method or scale. Below is the summary of the Cronbach's Alpha value for the variables include Perceived Ease to Use (PEU), Perceived Usefulness (PU), System Quality (SQ), Social Norm (SN) and Behavior Intention (BI).

Table 1. Summary of Cronbach's Alpha for Each Variable

| Variables | Number of Items | Items Dropped | Cronbach's Alpha, α |
|---|---|---|---|
| Perceived Ease to Use | 6 | - | 0.877 |
| Perceived Usefulness | 5 | - | 0.893 |
| System Quality | 6 | - | 0.852 |
| Social Norm | 4 | - | 0.821 |
| Behavioral Intention | 4 | - | 0.898 |

As shown in Table 1, all the variables have the positive value for Cronbach's Alpha and all are above 0.70. The reliability test on the items for each variable are consistent and inter-related to each other, hence no item is being dropped.

b) **Descriptive Analysis:** The descriptive analysis is used to summarize the sample and the measure. It is used to determine the level of user intention to adopt the proposed solution, i-Future. As in the questionnaire, the 7-Likerts Scale is used from the 1-Strongly Disgaree to 7-Strongly Agree. From the descriptive analysis perform using SPSS software, it shows that most of the respondent "agree" to adopt i-Future and the level of acceptance is more than 5. Table 2 shows the summary of descriptive analysis.

Table 2. Summary of Descriptive Analysis

|  |  | Mean_PEU | Mean_PU | Mean_SQ | Mean_SN | Mean_BI |
|---|---|---|---|---|---|---|
| N | Valid | 104 | 104 | 104 | 104 | 104 |
|  | Missing | 0 | 0 | 0 | 0 | 0 |
| Mean |  | 5.5705 | 5.4942 | 5.8253 | 5.4327 | 5.4038 |
| Std. Deviation |  | .74380 | .76168 | .73051 | .91855 | .94014 |
| Variance |  | .553 | .580 | .534 | .844 | .884 |
| Kurtosis |  | -.063 | .457 | -.079 | -.200 | .581 |
| Std. Error of Kurtosis |  | .469 | .469 | .469 | .469 | .469 |
| Range |  | 3.50 | 4.00 | 3.00 | 4.00 | 5.00 |
| Minimum |  | 3.50 | 3.00 | 4.00 | 3.00 | 2.00 |
| Maximum |  | 7.00 | 7.00 | 7.00 | 7.00 | 7.00 |

c) **Correlation Analysis:** The result shows that there is a significant relationship between Perceived Ease to Use (PU), Perceived Usefulness (PU), System Quality (SQ) and Social

Norm (SN) towards Behavioral Intention (BI). Table 3 shows the summary of correlation analysis.

Table 3. Summary of Correlation Analysis

**Correlations**

|  |  | Mean_PEU | Mean_PU | Mean_SQ | Mean_SN | Mean_BI |
|---|---|---|---|---|---|---|
| Mean_PEU | Pearson Correlation | 1 | .715** | .664** | .660** | .679** |
|  | Sig. (2-tailed) |  | .000 | .000 | .000 | .000 |
|  | N | 104 | 104 | 104 | 104 | 104 |
| Mean_PU | Pearson Correlation | .715** | 1 | .592** | .626** | .659** |
|  | Sig. (2-tailed) | .000 |  | .000 | .000 | .000 |
|  | N | 104 | 104 | 104 | 104 | 104 |
| Mean_SQ | Pearson Correlation | .664** | .592** | 1 | .699** | .578** |
|  | Sig. (2-tailed) | .000 | .000 |  | .000 | .000 |
|  | N | 104 | 104 | 104 | 104 | 104 |
| Mean_SN | Pearson Correlation | .660** | .626** | .699** | 1 | .668** |
|  | Sig. (2-tailed) | .000 | .000 | .000 |  | .000 |
|  | N | 104 | 104 | 104 | 104 | 104 |
| Mean_BI | Pearson Correlation | .679** | .659** | .578** | .668** | 1 |
|  | Sig. (2-tailed) | .000 | .000 | .000 | .000 |  |
|  | N | 104 | 104 | 104 | 104 | 104 |

**. Correlation is significant at the 0.01 level (2-tailed).

d) **Regression Analysis:** From the linear regression analysis result, it shows that H1, H2, H3, and H4 were accepted and there is a linear relationship between the Perceived Ease to Use (PEU), Perceived Usefulness (PU), System Qulaity (SQ), and Social Norm (SN) towards the dependent variable, which is the Behavioral Intention (BI). After identifying the linear regression, the multiple regressions are computed using SPSS. The result of multiple regressions is shown in the Table 4 and 5.

Table 4. Summary of Linear Regression

| Hypothesis | Beta (β) | R Square | Significant Level, p | Conclusion |
|---|---|---|---|---|
| H1 | 0.679 | 0.461 | **0.000 | Hypothesis Accepted |
| H2 | 0.659 | 0.434 | **0.000 | Hypothesis Accepted |
| H3 | 0.578 | 0.334 | **0.000 | Hypothesis Accepted |
| H4 | 0.668 | 0.447 | **0.000 | Hypothesis Accepted |

Table 5. Summary of Multiple Regressions

| Variable | Dependent = Behavioral Intention Standardized Coefficients (β) |
|---|---|
| Perceived Ease to Use (PEU) | 0.276** |
| Perceived Usefulness (PU) | 0.250** |
| Social Norm (SN) | 0.307** |
| R | 0.759 |
| R Square | 0.576 |
| Adjusted R Square | 0.559 |
| Significance, p | 0.000 |
| Durbin-Watson | 1.850 |

\** Significant at p<0.01

Based on the multiple regressions analysis, it shows that the multiple independent variables include Perceived Ease to Use (PEU), Perceived Usefulness (PU), Social Norm (SN) has the significant impact to the dependent variable, Behavioral Intention (BI). This has proof the relationship of H5 in the proposed research model.

## 7. Discussion

Based on the data analysis, the research questions are being answered. From the finding, the users are "agree" with the proposed solution and and their intention level to use the system is more than five. Besides, through the result on the finding, H1, H2, H3 and H4 are accepted. However, when compute the multiple regressions, the result shows that only PEU, PU and SN have the significance towards the Behavioral Intention (BI) whereas System Quality (SQ) shows little or no significance.

## 8. Conclusion

i-Future is a web-based solution specially designed to solve the limitation in current existing educational system with the aim to create more critical thinking intellectuals to better prepare them in the changing world and ensure the world class talent based for a higher income nation. Through the immersive learning environment provided in i-Future, the children can "enter to learn, leave to achieve". As in the finding of the research, the parents in Malaysia are agree with proposed solution and intend to adopt the system. Hence, the beta version of i-Future will be developed based on the feedback collected such as PEU, PU, and SN has the positive significant towards BI.

## 9. Acknowledgement

Special Thanks to Mr. Mohammad Ali Bagheri for his help and contribution in preparing and publishing this paper. Moreover, the authors would like to thank Universiti Sains Malaysia (USM) as this research has been supported from the Research University Grant (RUI) [Account Number: 1001/PKOMP/811251] and from the Short Term Research Grant [Account Number: 304/PKOMP/6312103] from the Universiti Sains Malaysia..

## References

[1] Malaysia Achieving the Millennium Devlopment Goals (Success and Challenges). Kuala Lumpur: United Nations Country Team Malaysia; 2005.

[2] Charles B. Learning Begins at Birth (The Earliest Years Count). Sate Early Childhood Policy Technical Assistance Network; 2005.

[3] Shore R. Rethinking the Brain: Early Childhood Brain Development. Families and Work Institute; 1997.

[4] Wahab A. World Class Education? Malaysia Ranked 52nd Again In Global Education Ratings. Retrieved on 13 May 2015 URL: http://www.malaysiandigest.com/news/553456-world-class-education-malaysia-ranked-52nd-again-in-global-education-ratings.html

[5] Leidner D E., Fuller M A. Improving Student Processing and Assimilation of Conceptual Information: GSS-Supported Collaborative Learning vs. Individual Constructive Learning. Proceedings of the 29th IEEE Annual Hawaii International Conference on System Sciences, 1996; 8-9.

[6] Tenth Malaysia Plan 2011-2015. Prime Minister's Department, Putrajaya. The Economic Planning Unit; 2010.

[7] Dewey J. Experience & Education, 1997. URL: Amazon.com.

[8] Davis F. User Acceptence of Information Technology: System Characteristics. User Perceptions and Behavior Impacts, 1993; 38; 11-12.

[9] Chien H. Consumer Adoption of e-Service: Integrating Technology Readiness with the Technology Acceptance Model. Technology Management: A Unifying Descripline for Melting the Boundaries, 2005; 4-5.

[10] Ajzen I. The Theory of Planned Behavior, Organizational Behavior and Human Decision Process, 1991; 50; 31-32.

[11] Karaiskos D. Understanding the Adoption of Mobile Data Services: Differences Among Mobile Portal and Mobile Internet Users, 2009; 4-6.

[12] DeLone W H., McLean E R. Information Systems Success: The Quest for the Dependent Variable. Information Systems Research, 1992; 37.